\documentclass[onecolumn,showpacs,preprintnumbers,amsmath,amssymb]{revtex4}

\usepackage{amsmath}    % need for subequations
\usepackage{graphicx}   % for figures
\usepackage{graphics}% Include figure files
\usepackage{dcolumn}% Align table columns on decimals point
\usepackage{bm}% bold math
\usepackage{color}

\usepackage{latexsym}
\usepackage{amssymb}
\usepackage{amsmath}

\begin{document}
\title{Multi-shell contribution to the Dzyaloshinskii--Moriya spiralling in MnSi-type crystals}
\author{Viacheslav A. Chizhikov\footnote{email: chizhikov@crys.ras.ru} and Vladimir E. Dmitrienko\footnote{email: dmitrien@crys.ras.ru}}
\affiliation{A.V.~Shubnikov Institute of Crystallography, 119333 Moscow, Russia}


\begin{abstract}%\Large
The transition from the microscopic Heisenberg model to the
macroscopic elastic theory is carried out for the chiral magnetics
of MnSi-type with the $B20$ crystal structure. Both exchange and
Dzyaloshinskii--Moriya (DM) interactions are taken into account
for the first, second, and third magnetic neighbors. The
particular components of the DM vectors of bonds are found, which
are responsible for (i) the global magnetic twist and (ii) the
canting between four different spin sublattices. A possible
mechanism for effective reinforcement of the global magnetic twist
is suggested: it is demonstrated that the components of the DM
vectors normal to corresponding interatomic bonds become very
important for the twisting power. The
Ruderman--Kittel--Kasuya--Yosida (RKKY) theory is used for model
calculation of exchange parameters.  It is found that just the
interplay between the exchange parameters of several magnetic
shells rather than the signs of DM vectors can be responsible for
the concentration-induced reverse of the magnetic chirality
recently observed in the Mn$_{1-x}$Fe$_{x}$Ge crystals.
\end{abstract}
\maketitle

%\Large

\section{Introduction}
\label{sec:intro}

Chiral spin textures are studied now very actively for possible
spin self-organization, unusual quantum transport phenomena and
spintronic applications. A well established mechanism of spin
chirality is the spin-orbit Dzyaloshinskii--Moriya (DM)
interaction which is responsible for intricate magnetic patterns
in the MnSi-type crystals. Even half a century after the discovery
of the strange magnetic properties of MnSi
\cite{Williams1966,Shinoda1966}, the magnetics with the $B20$
crystal structure still amaze us with the variety and complexity
of their magnetic phases and electronic properties
\cite{Stishov2011} contrasting the simplicity of the crystalline
arrangement (only four magnetic atoms per a unit cell). Among the
magnetic phases, both experimentally observed and hypothetical,
are simple and cone helices \cite{Ishikawa1976,Motoya1976}, the
Skyrmions and their lattices associated with the recently found
$A$-phase \cite{Rossler2006,Grigoriev2006a,Munzer2010,Adams2011},
possible 3D structures \cite{Tewari2006,Binz2006,Hamann2011}
similar to the blue phases of liquid crystals, etc. This variety
is due to, first, the lack of inverse and mirror symmetries, which
gives rise to the chirality of the crystalline and spin
structures; second, the frustrations
\cite{Hopkinson2006,Chizhikov2012} resulting from nontrivial
topology of the trillium lattice, that introduces a competition of
various interactions between different pairs of atoms.

Beginning from the discovery of the chiral magnetic properties of
MnSi in 1976 \cite{Ishikawa1976,Motoya1976} till present day the
most used approach to describe and predict twisted magnetic
structures remains the phenomenological theory based on the
Ginsburg--Landau free energy with an additional term first
introduced by Dzyaloshinskii
\cite{Dzyaloshinskii58,Bak80,Nakanishi1980,Tewari2006,Binz2006}.
However, the approach, which uses our knowledge of the system
symmetry, is not able to say anything about the values of
coefficients in the free energy, for instance, how they are
connected with the real interactions between atoms.

The microscopic theories, e.g. the model of the classical
Heisenberg ferromagnetics with a spin-orbit term originally
developed by Moriya \cite{Moriya60b}, have in their turn the
shortcoming that the number of variables, including the spins of
all the magnetic atoms, is infinite, and therefore they are
difficult to use for any analytical computations. Nevertheless,
in spite of some doubts about validity the Heisenberg model in
the itinerant magnetics, it is often used for digital simulations
\cite{Hamann2011,Hopkinson2009}. For this model it is of
great importance to know the parameters of different interactions
between pairs of spins: $J_{ij}$ for exchange coupling of $i$th
and $j$th atoms, and vectors $\mathbf{D}_{ij}$ for DM
interactions. Those parameters can be obtained in two ways: from
comparison of theoretical results with an experimental data and by
means of {\it ab intio} calculations.

\begin{figure}[h]
\includegraphics[width=7cm]{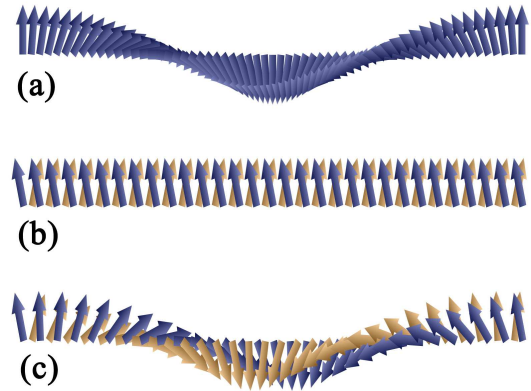}
\caption{\label{fig-canting} (color online). (a) The 1D twisted
magnetic structure; the spins rotate in the plane perpendicular to
the propagation vector. (b) The canting in the absence of a twist;
the lattice is divided into two sublattices (blue and
sand-coloured arrows) with the spins tilted by a constant angle.
(c) The canting and the twist; the canting of the sublattices
leads to the combination of two helices with the same propagation
vector and a constant phase shift.}
\end{figure}

In our recent work \cite{Chizhikov2012}, using a coarse-grained
approximation, the phenomenological constants ${\cal J}$ and
${\cal D}$ of the elastic free energy have been connected with
the corresponding parameters of the microscopic theory. A
surprising detail was the recognition of inter-sublattice canting
as a new microscopic feature of the magnetic structures in the
MnSi-type crystals. Fig.~{\ref{fig-canting}} shows the difference
between the twist and the canting by example of a 1D spin chain.
An experimental confirmation of the canting
could give us an argument for applicability of the Heisenberg
model to the itinerant ferromagnetics of MnSi-type
\cite{Dmitrienko2012}. Similar features could be observed in
dielectric chiral magnetics like recently studied multiferroics
Cu$_2$OSeO$_3$ \cite{Seki2012,Adams2012,GongXiang2012}, BiFeO$_3$
\cite{Zvezdin2012} and
langasite-type crystals \cite{Pikin2012}.

Though the theory with only nearest neighbors interactions is
frequently used to simulate spin structures
\cite{Hopkinson2009,Hamann2011,Chizhikov2012}, it does not always
give an adequate physical description. Indeed, very often the
coupling between second and next neighbors is comparable or even
more considerable than that of the nearest neighbors. Thus, for
example, in the weak ferromagnetic $\alpha$-Fe$_2$O$_3$, the
antiferromagnetic exchange interactions and DM interactions are
expected to be the most strong for third and fourth neighbors (see
{\it ab initio} calculations in \cite{Mazurenko2005}); this is
also supported by the experimental data for the Heisenberg
exchange parameters \cite{Samuelsen1970}.

It also could happen that only with the
interactions between non-nearest neighbors one
can describe an experimentally observed
phenomenon. For example, if in a 1D spin chain
with the ferromagnetic coupling between the
nearest neighbors ($J_1 > 0$) one turns on the
antiferromagnetic interaction with the second
neighbors ($J_2 < 0$), then for $|J_2| > J_1/4$ a
spiral magnetic structure appears owing to
spontaneous breaking of the inverse symmetry.
Such approach can explain the phase transition
appearing at $\sim$28 K and inducing
ferroelectricity in the multiferroic
Tb(Dy)MnO$_3$ crystal
\cite{Kenzelmann2005,Cheong2007}. The approach
works also in the itinerant magnetics of
MnSi-type. Just the competition between
ferromagnetic coupling of the nearest spins ($J_1
> 0$) and antiferromagnetic coupling of the
second and third neighbors ($J_2 \approx J_3 <
0$) was utilized to explain the observed
alignment of magnetic helices along
crystallographic directions $\langle 110 \rangle$
in MnSi at high pressure \cite{Hopkinson2007}.

In this paper, we suggest a possible mechanism for effective
reinforcement of the twist terms in the chiral spin structures of
the $B20$ magnetics, taking into account the interactions with
non-nearest neighbors. The reinforcement is caused by an
interplay between the exchange coupling with the second and
third neighbors and the canting from DM interactions. In
Sections \ref{sec:interactions}--\ref{sec:energy} the transition
from the Heisenberg microscopic model to the continuous
phenomenological one is performed, and the spin-orbit terms in the
phenomenological energy are found up to the contributions of the
order of $(D/J)^2$. In Sec.~\ref{sec:extratwist} the possibility
of an extra twist induced by the canting is demonstrated. In
Sec.~\ref{sec:rkky} the exchange interaction parameters are
estimated within the RKKY theory. In Sec.~\ref{sec:experimental}
the possibility of an experimental proof of the canting existence
is discussed. The possible applications of the theory are
suggested in Sec.~\ref{sec:discussion}.

\section{Dzyaloshinskii--Moriya interaction in microscopical and phenomenological approaches}
\label{sec:interactions}

Both phenomenological and microscopical theories
describing twisted MnSi-type magnetics contain  terms
induced by chirality of the system and associated
with the Dzyaloshinskii--Moriya interaction of
the spin-orbit nature. In the  Heisenberg model,
the extra term being associated with an
individual bond, e.g. connecting magnetic atoms 1
and 2, can be expressed as
\begin{equation}
\label{eq:heisenberg}
\mathbf{D}_{12} \cdot [\mathbf{s}_1 \times \mathbf{s}_2] ,
\end{equation}
where $\mathbf{s}_1$ and $\mathbf{s}_2$ are the classical spins of
the atoms, $\mathbf{D}_{12}$ is the DM vector of the bond 1-2. In
principle, there could be magnetic moments at Si atoms
\cite{Brown} but this effect will be neglected below. Let us
shortly describe the properties of the DM vector \cite{Moriya60b}.
(i) As it is obvious from Eq.~(\ref{eq:heisenberg}), the sign of
the DM vector depends on which atom of the bond we consider as the
first. Indeed, because the cross product changes its sign when
$\mathbf{s}_1$ and $\mathbf{s}_2$ are rearranged, then
$\mathbf{D}_{21} = -\mathbf{D}_{12}$. (ii) The structure changes
its chirality under inversion, whereas the energy remains
unchanged, which means that $\mathbf{D}_{12}$ is a pseudovector,
because $[\mathbf{s}_1 \times \mathbf{s}_2]$ is a pseudovector as well.
(iii) The DM vector of a
bond possesses the local symmetry of the bond. Thus, in MnSi the
DM vectors are of their most general form, i.e. they have three
independent components, because the Mn-Mn bonds in the $B20$
structure do not possess any internal symmetry element. (iv) The
DM vectors vary from bond to bond; the DM vectors of the
equivalent bonds have the same length but may be of different
orientation (just the case of MnSi). In particular, if two bonds
are connected by the rotation symmetry transformation of the space
group $P2_13$ of the crystal, then their DM vectors do by the
corresponding rotation of the point group $23$.

In the phenomenological theory, the chiral
interaction is induced by the following extra
term
\begin{equation}
\label{eq:phenomenological}
{\cal D} \mathbf{M} \cdot [\boldsymbol{\nabla} \times \mathbf{M}]
\end{equation}
in the expression for the magnetic energy
density. Here $\mathbf{M}$ is a continuous field
of the magnetic moment, ${\cal D}$ is a
pseudoscalar constant of the interaction.

Because both the theories describe the same
matter, there should be a relationship between
them. In particular, the terms
(\ref{eq:heisenberg}) and
(\ref{eq:phenomenological}) of the different
approaches should be somehow connected. Indeed,
it was shown in Ref.~\cite{Chizhikov2012} that in
the nearest neighbors approximation the constant
${\cal D}$ of the phenomenological theory is
proportional to the component $(D_x-2D_y-D_z)$ of
the DM vector of the bond $(-2x, \frac{1}{2},
\frac{1}{2}-2x)$ between neighboring manganese
atoms.

However, there is a problem here. According to
different spin-orbit calculation schemes
\cite{Mazurenko2005,Cheong2007,Moskvin1977,Shekhtman93,Sergienko2006,Katsnelson10},
the DM vector associated with a bond should be
perpendicular or almost perpendicular to the
bond. In the present case it means that the
component $(D_x-2D_y-D_z)$, which lies almost
along the bond, constitutes only a small part of
the DM vector length. Taking into account that
having the relativistic nature spin-orbit DM
interaction serves as a small additive to the
ferromagnetic exchange coupling, we can conclude
that the twist observed in the MnSi-type
crystals, particularly in MnGe, seems to be
abnormally strong. A possible solution of this
problem is that the nearest neighbors
approximation is not sufficient and hereinafter
in the paper we develop this idea in details.

In the following sections we will show, how to
perform transformation from the microscopic
Heisenberg model to the macroscopic
elastic theory.

\section{From the Heisenberg model to a continuous approximation}
\label{sec:continuous}

In the classical Heisenberg model of magnetics
with an extra interaction of the
Dzyaloshinskii--Moriya type, the energy of the
system is expressed as a sum of pair interactions
between magnetic atoms and the interaction of
individual atoms with an external magnetic field
$\mathbf{H}$:
\begin{equation}
\label{eq:energy1}
E = \sum_{cells} \left( \sum_{n,\{ij\}} \left\{ -\tilde{J}_n \mathbf{s}_i \cdot \mathbf{s}_j + \mathbf{D}_{n,ij} \cdot [\mathbf{s}_i \times \mathbf{s}_j] - \frac{(\mathbf{D}_{n,ij} \cdot \mathbf{s}_i)(\mathbf{D}_{n,ij} \cdot \mathbf{s}_j)}{2J_n} \right\} - g \mu_B \mathbf{H} \cdot \sum_{i=1}^4 \mathbf{s}_i \right) .
\end{equation}
Here the external summation is over all the unit
cells of the crystal, $n$ enumerates magnetic
coordination shells, the sum $\{ij\}$ is over all
the bonds of the shell $n$, the sum over $i$ is
taken over all magnetic atoms in the unit cell;
$\tilde{J}_n = J_n - (D_n^2/4J_n)$, $J_n$ is the
exchange coupling interaction parameter of the
$n$th shell, $\mathbf{D}_{n,ij}$ is the DM vector
of the bond $ij$ of the $n$th shell,
$\mathbf{s}_i$ are classical spins of magnetic
atoms, $|\mathbf{s}|=1$, $g\mu_B$ is the effective
magnetic moment of the atom. The terms of the
order of $D_n^2$ are usually ignored but
sometimes they can be important
\cite{Shekhtman,Yildirim}.

The MnSi-type crystal has $B20$ structure
($P2_13$ space group) with four magnetic (Mn)
atoms occupying crystallographically equivalent
$4a$ positions within a unit cell, $\mathbf{r}_1
= ( x, x, x)$, $\mathbf{r}_2 = ( \frac{1}{2}-x, 1-x, \frac{1}{2}+x)$, $\mathbf{r}_3 = ( 1-x,
\frac{1}{2}+x, \frac{1}{2}-x)$,
$\mathbf{r}_4 = ( \frac{1}{2}+x, \frac{1}{2}-x,
1-x)$ \cite{tables}. The shortest bonds between
the magnetic atoms have the length
$\sqrt{\frac{1}{2} - 2x  + 8x^2}$. The next
environment consists of two close magnetic shells
with the radii $\sqrt{\frac{3}{2} - 6x  + 8x^2}$
and $\sqrt{\frac{1}{2} + 2x  + 8x^2}$. The
parameter $x$ defines which of them is closer
to the initial atom. When $x > \frac{1}{8}$
($x\approx 0.138$ in the case of MnSi), then the
shell with the radius $\sqrt{\frac{3}{2} - 6x  +
8x^2}$ is closer, so we will refer to it as the
second one. When $x = \frac{1}{8}$, the 2nd and
3rd neighbors are at the same distance so that
the manganese sublattice gains the space group
$P4_332$ \cite{Dmitriev2012}, connecting the 2nd and 3rd shells by a
symmetry transformation. This could establish a
linkage between the DM vectors of the shells, but
the silicon sublattice (having in its turn the
space group $P4_132$, when $x_{Si} =
\frac{7}{8}$) breaks the symmetry and the DM
vectors of the shells are different even in this
case. If $x < \frac{1}{8}$, then the 3rd
neighbors become closer than the 2nd ones.

\begin{figure}[h]
\includegraphics[width=7cm]{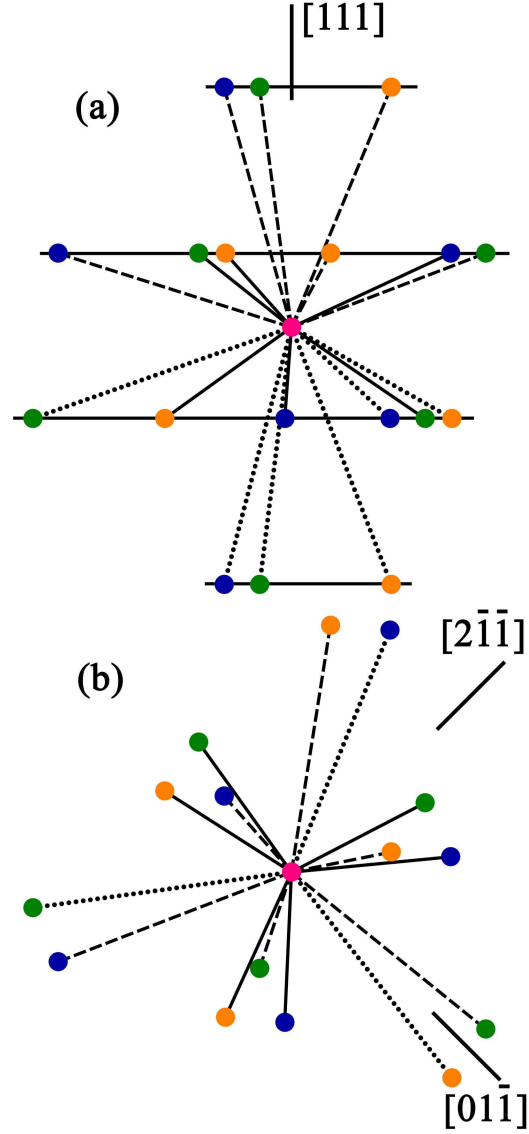}
\caption{\label{fig-neighbors} (color online). Nearest magnetic
neighborhood of a manganese atom in MnSi. The central atom
(magenta) belongs to the first manganese sublattice, its neighbors
are in 2nd (green), 3rd (orange) and 4th (blue) ones. The shortest
bonds are given by solid lines, the bonds with the atoms from the
2nd and 3rd magnetic shells are shown by dashes and dots,
correspondingly. (a) The 3-fold axis $[111]$ is vertical. The
horizontal lines designate equidistant atomic planes containing
the manganese atoms of 2nd, 3rd and 4th sublattices. (b) The
projection on the plane perpendicular to the axis $[111]$. Two
atomic triangles from (a), the upper and the lowest, belonging to
different shells are situated just one above the other.}
\end{figure}

Every magnetic atom possesses six neighbors in
each of its three nearest magnetic shells, so
there are 36 bonds in a unit cell (in twelves for
each kind). Fig.~\ref{fig-neighbors} shows the
magnetic environment of a manganese atom in MnSi.
The atom lies on a 3-fold axis. Each of the first
three magnetic shells consists of two equilateral
triangles. All the atoms of the 2nd shell are on
one side of the plane perpendicular to the 3-fold
axis, all the atoms of the 3rd shell are situated
on the other side of the plane.

Let $( D_{1x}, D_{1y}, D_{1z})$ be the
coordinates of the vector $\mathbf{D}_{1,13}$
corresponding to the bond $\mathbf{b}_{1,13} = (
-2x, \frac{1}{2}, \frac{1}{2}-2x)$ directed from
$\mathbf{r}_1$ to $\mathbf{r}_3-(1,0,0)$, $(
D_{2x}, D_{2y}, D_{2z})$ be the coordinates of
the vector $\mathbf{D}_{2,13}$ corresponding to
the bond $\mathbf{b}_{2,13} = ( 1-2x,
\frac{1}{2}, \frac{1}{2}-2x)$ directed from
$\mathbf{r}_1$ to $\mathbf{r}_3$, and $( D_{3x},
D_{3y}, D_{3z})$ be the coordinates of the vector
$\mathbf{D}_{3,13}$ corresponding to the bond
$\mathbf{b}_{3,13} = ( -2x, \frac{1}{2},
-\frac{1}{2}-2x)$ directed from $\mathbf{r}_1$ to
$\mathbf{r}_3-(1,0,1)$.

We specify directions for all twelve bonds of
a shell in the way that all directed distances
$\mathbf{b}_{ij}$ between neighboring atoms
connect to each other by the symmetry
transformations of the point group $23$, so
corresponding DM vectors are connected by the
same transformations. In the MnSi-type crystals
there is only one type of magnetic bonds directed
from a central atom of the type $t$ to an atom of
the type $t^\prime \neq t$ at a given shell ($n =
1, 2, 3$). We can take advantage from this fact
and introduce for each pair $ij$ ($i, j = 1, 2,
3, 4$, $i \neq j$) a local triad
\begin{equation}
\label{eq:triad}
( \boldsymbol{\tau}_i - \boldsymbol{\tau}_j ) \perp ( \boldsymbol{\tau}_i + \boldsymbol{\tau}_j ) \perp [\boldsymbol{\tau}_i \times \boldsymbol{\tau}_j] ,
\end{equation}
and then use it as a basis for vectors
$\mathbf{b}_{ij}$ and $\mathbf{D}_{ij}$. Here
$\boldsymbol{\tau}_i$ is a vector directed along
the 3-fold axis passing through the position
$\mathbf{r}_i$:
\begin{equation}
\label{eq:tau}
\begin{array}{l}
\boldsymbol{\tau}_1 = ( 1, 1, 1) , \\
\boldsymbol{\tau}_2 = ( -1, -1, 1) , \\
\boldsymbol{\tau}_3 = ( -1, 1, -1) , \\
\boldsymbol{\tau}_4 = ( 1, -1, -1) .
\end{array}
\end{equation}
Then the bond vector and the DM vector of an
arbitrary bond in three first magnetic shells can
be written as
\begin{subequations}     % need amsmath package
\label{eq:bij}
\begin{eqnarray}
\mathbf{b}_{1,ij} & = & (-8x+1) (\boldsymbol{\tau}_i - \boldsymbol{\tau}_j) / 8 + (\boldsymbol{\tau}_i + \boldsymbol{\tau}_j) / 4 + [\boldsymbol{\tau}_i \times \boldsymbol{\tau}_j] / 8 , \\
\mathbf{b}_{2,ij} & = & (-8x+3) (\boldsymbol{\tau}_i - \boldsymbol{\tau}_j) / 8 + (\boldsymbol{\tau}_i + \boldsymbol{\tau}_j) / 4 - [\boldsymbol{\tau}_i \times \boldsymbol{\tau}_j] / 8 , \\
\mathbf{b}_{3,ij} & = & (-8x-1) (\boldsymbol{\tau}_i - \boldsymbol{\tau}_j) / 8 + (\boldsymbol{\tau}_i + \boldsymbol{\tau}_j) / 4 - [\boldsymbol{\tau}_i \times \boldsymbol{\tau}_j] / 8 ,
\end{eqnarray}
\end{subequations}
\begin{equation}
\label{eq:Dij}
\mathbf{D}_{n,ij} = \frac{D_{n+}}{4} (\boldsymbol{\tau}_i - \boldsymbol{\tau}_j) + \frac{D_{ny}}{2} (\boldsymbol{\tau}_i + \boldsymbol{\tau}_j) - \frac{D_{n-}}{4} [\boldsymbol{\tau}_i \times \boldsymbol{\tau}_j] ,
\end{equation}
with $D_{n\pm} = D_{nx} \pm D_{nz}$; here
$\mathbf{D}_{n,ji} \neq -\mathbf{D}_{n,ij}$
because the indices $ij$ and $ji$ designate two
different bonds.

\begin{figure}[h]
\includegraphics[width=7cm]{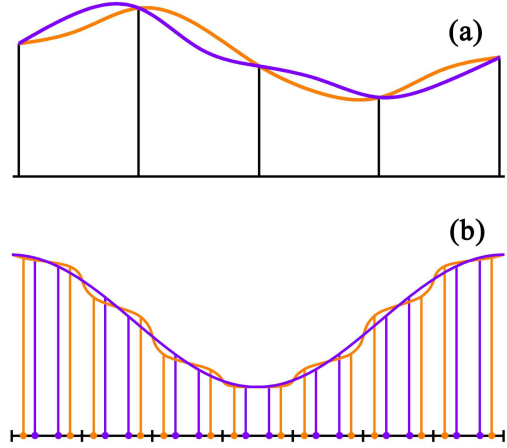}
\caption{\label{fig-xid} (color online). (a) The ambiguity of the
transition from discrete to continuous model: two different smooth
functions having the equal values in a discrete set of points. (b)
The gain from the arbitrary choice of the parameter $x$. The
initial curve (orange) is plotted with the use of the function
values given in the discrete set of points with the coordinates
$x$ and $\bar{x}$ within the cells of a 1D crystal. When the
points are shifted to the center of the cells without a change of
function values, the view of the curve changes (blue).}
\end{figure}

Notice that the energy (\ref{eq:energy1}) does
not depend on the parameter $x$, so it can be
chosen arbitrarily, making convenient transition
to the continuous approximation
(Fig.~\ref{fig-xid}(b)). Indeed, to provide the
transition the continuous unimodular vector
functions will be defined, $\hat{\mathbf{s}}_1$,
$\hat{\mathbf{s}}_2$, $\hat{\mathbf{s}}_3$,
$\hat{\mathbf{s}}_4$, which coincide in special
points with the spin values in corresponding
atomic positions. If the parameter $x$ is equal
to its experimental value, e.g. $x=0.138$ in
MnSi, then we should take the functions in the
real atomic positions in order to obtain the real
spins. In the case of an arbitrary choice of $x$
we should take the function values in the points
shifted from the real atomic positions.

It is convenient to choose
the parameter $x$ from the condition
\begin{equation}
\label{eq:x}
\sum_{n,\{ij\}} \tilde{J}_n (\boldsymbol{\tau}_{i}-\boldsymbol{\tau}_{j}) \otimes \mathbf{b}_{n,ij} = 0 ,
\end{equation}
where $\otimes$ means the direct product of two vectors. Notice
that this $3 \times 3$-tensor is proportional to the unit one due
to the averaging over the symmetry elements of the point group
$23$ (see Appendix, Eq.~(\ref{eq:aijbij})).

Thus, for example, in the nearest neighbors
approximation the condition gives $x =
\frac{1}{8}$, which is shown in
Ref.~\cite{Chizhikov2012} to be necessary in
order to obtain smooth spin functions.

For $n = 1, 2, 3$ using Eqs.~(\ref{eq:sumtaub})
we find from Eq.~(\ref{eq:x})
\begin{equation}
\label{eq:xexch}
x_{exch} = \frac{\tilde{J}_1 + 3 \tilde{J}_2 - \tilde{J}_3}{8(\tilde{J}_1 + \tilde{J}_2 + \tilde{J}_3)} ,
\end{equation}
where the index ``exch'' means that $x_{exch}$ is not a real
coordinate, but some physical parameter expressed trough the
exchange interaction constants. In untwisted spin structures,
the canting is determined wholly by the DM interactions. In
twisted states, an extra canting arises induced by a
disagreement of phases of the helices connected with different
magnetic sublattices. The physical meaning of the ``exchange''
coordinates is that the spin shift to the new positions removes
the canting appearing due to the spiralling.

The bond $ij$ of the $n$th shell connects the
function values $\hat{\mathbf{s}}_i(\mathbf{r})$
and $\hat{\mathbf{s}}_j(\mathbf{r} +
\mathbf{b}_{n,ij})$. Using continuity of the
functions $\hat{\mathbf{s}}$ we can write
\begin{equation}
\label{eq:taylor}
\hat{\mathbf{s}}_j (\mathbf{r} + \mathbf{b}_{n,ij}) = {\cal B}_{n,ij} \hat{\mathbf{s}}_j (\mathbf{r}) ,
\end{equation}
where
\begin{equation}
\label{eq:calB}
{\cal B}_{n,ij} \equiv \exp(\mathbf{b}_{n,ij} \cdot \boldsymbol{\nabla}) = 1 + (\mathbf{b}_{n,ij} \cdot \boldsymbol{\nabla}) + \frac12 (\mathbf{b}_{n,ij} \cdot \boldsymbol{\nabla})^2 + \ldots
\end{equation}
is the operator representing the Taylor series expansion, and pass
from the summation over the unit cells to the integration over the
crystal volume (in the chosen units, the lattice parameter
$a=1$ and the unit cell volume is equal to 1):
\begin{equation}
\label{eq:energy2}
\begin{array}{ll}
E = & \int_{V} {\cal E} d\mathbf{r} ,
\end{array}
\end{equation}
\begin{equation}
\label{eq:energy2b}
\begin{array}{ll}
{\cal E} = & \sum_{n,\{ij\}} \{ -\tilde{J}_n \hat{\mathbf{s}}_i \cdot {\cal B}_{n,ij} \hat{\mathbf{s}}_j + \mathbf{D}_{n,ij} \cdot [\hat{\mathbf{s}}_i \times {\cal B}_{n,ij} \hat{\mathbf{s}}_j] \\
 & - (\mathbf{D}_{n,ij} \cdot \hat{\mathbf{s}}_i)(\mathbf{D}_{n,ij} \cdot {\cal B}_{n,ij} \hat{\mathbf{s}}_j)/(2J_n) \} - g \mu_B \mathbf{H} \cdot \sum_{i=1}^4 \hat{\mathbf{s}}_i .
\end{array}
\end{equation}

\section{Magnetic moment density, calculus of variations and perturbation theory}
\label{sec:moment}

In the phenomenological theory, the magnetic moment density
$\mathbf{M}$ is used as an order parameter and considered as an
experimentally observed physical quantity. Nevertheless, when we
try to express a smooth continuous function $\mathbf{M}$ through a
discrete distribution of spins of the magnetic atoms, an ambiguity
arises from the fact that we can determine the weight function in
different ways when averaging the spins on a local volume. In this
case the ambiguity results in that the smooth functions
$\hat{\mathbf{s}}_i$ having specified values in a discrete set of
points can be defined by infinite number of ways
(Fig.~\ref{fig-xid}(a)). We will avoid the problem, supposing that
the functions $\hat{\mathbf{s}}_i$ are defined in the most
convenient way.

It would be natural to define the magnetization as
\begin{equation}
\label{eq:momentum}
\mathbf{M} = 4 g \mu_B \mathbf{m} ,
\end{equation}
\begin{equation}
\label{eq:momentumb}
\mathbf{m} = \frac14 \sum_{i=1}^4 \hat{\mathbf{s}}_i .
\end{equation}
In analogy with Eq.~(\ref{eq:momentumb}) we introduce the canting tensor
\begin{equation}
\label{eq:udef2}
u_{\sigma\alpha} = \frac14 \sum_{i=1}^4 \tau_{i\sigma} s_{i\alpha} , \phantom{x} \alpha, \sigma = x, y, z ,
\end{equation}
so that
\begin{equation}
\label{eq:udef}
\hat{s}_{i\alpha} = m_\alpha + \tau_{i\sigma} u_{\sigma\alpha} .
\end{equation}
Hereinafter the summation on repeated greek indices is
implied. Four conditions $|\hat{\mathbf{s}}_i| =
1, i=1,2,3,4$, can be rewritten using the
invariants
\begin{equation}
\label{eq:invariantI0}
I_0 \equiv \frac14 \sum_{i=1}^4 \hat{s}_i^2 = m^2 + u_{\sigma\alpha} u_{\sigma\alpha} = 1 ,
\end{equation}
\begin{equation}
\label{eq:invariantsI}
I_\sigma \equiv \frac14 \sum_{i=1}^4 \tau_{i\sigma} \hat{s}_i^2 = 2 m_\alpha u_{\sigma\alpha} + |\varepsilon_{\sigma\beta\gamma}| u_{\beta\alpha} u_{\gamma\alpha} = 0 .
\end{equation}

In order to express the energy as a functional of the magnetic
moment $\mathbf{m}$, the energy should be minimized by the canting
tensor components using calculus of variations. The variation of
the energy is
\begin{equation}
\label{eq:varenergy}
\delta\tilde{E} = \int_{V} \delta\tilde{\cal E} d\mathbf{r} = \int_{V} \delta u_{\sigma\alpha} \Psi_{\sigma\alpha} d\mathbf{r} ,
\end{equation}
where the integrand now includes the Lagrange terms:
\begin{equation}
\label{eq:varenergyb}
\tilde{\cal E} = {\cal E} + \lambda_0 I_0 + \lambda_x I_x + \lambda_y I_y + \lambda_z I_z .
\end{equation}
Taking into account Eq.~(\ref{eq:udef}),
\begin{equation}
\label{eq:varcale}
\begin{array}{rl}
\Psi_{\sigma\alpha} = & \sum_{n,\{ij\}} \{ -\tilde{J_n} [ \tau_{i\sigma} {\cal B}_{n,ij} + \tau_{j\sigma} {\cal B}_{n,ij}^{-1} ] m_\alpha \\
 & -\tilde{J}_n [ \tau_{i\sigma} \tau_{j\rho} {\cal B}_{n,ij} + \tau_{j\sigma} \tau_{i\rho} {\cal B}_{n,ij}^{-1} ] u_{\rho\alpha} \\
 & -\varepsilon_{\alpha\beta\gamma} D_{n,ij\beta} [ \tau_{i\sigma} {\cal B}_{n,ij} - \tau_{j\sigma} {\cal B}_{n,ij}^{-1} ] m_\gamma \\
 & -\varepsilon_{\alpha\beta\gamma} D_{n,ij\beta} [ \tau_{i\sigma} \tau_{j\rho} {\cal B}_{n,ij} - \tau_{j\sigma} \tau_{i\rho} {\cal B}_{n,ij}^{-1} ] u_{\rho\gamma} \\
 & - (1/2J_n) {D}_{n,ij\alpha} {D}_{n,ij\beta} [ \tau_{i\sigma} {\cal B}_{n,ij} + \tau_{j\sigma} {\cal B}_{n,ij}^{-1} ] m_\beta \\
 & - (1/2J_n) {D}_{n,ij\alpha} {D}_{n,ij\beta} [ \tau_{i\sigma} \tau_{j\rho} {\cal B}_{n,ij} + \tau_{j\sigma} \tau_{i\rho} {\cal B}_{n,ij}^{-1} ] u_{\rho\beta} \} \\
 & + 2 \lambda_0 u_{\sigma\alpha} + 2 \lambda_\sigma m_\alpha + 2 |\varepsilon_{\sigma\beta\gamma}| \lambda_\beta u_{\gamma\alpha} .
\end{array}
\end{equation}
Here
\begin{equation}
\label{eq:calB2}
{\cal B}_{n,ij}^{-1} = \exp(-\mathbf{b}_{n,ij} \cdot \boldsymbol{\nabla})
\end{equation}
and the evident equation was used, which corresponds to the integration in infinite volume,
\begin{equation}
\label{eq:parts}
\int f(\mathbf{r}) {\cal B} g(\mathbf{r}) d\mathbf{r} = \int g(\mathbf{r}) {\cal B}^{-1} f(\mathbf{r}) d\mathbf{r} .
\end{equation}

The minimum of the energy is determined from the condition
$\delta\tilde{\cal E} = 0$ with arbitrary functions $\delta
u_{\sigma\alpha}$, so the problem is reduced to the system of 9
equations
\begin{equation}
\label{eq:Psi}
\Psi_{\sigma\alpha} = 0 , \phantom{x} \sigma, \alpha = x, y, z ,
\end{equation}
with the extra conditions (\ref{eq:invariantI0}),
(\ref{eq:invariantsI}) determining the functions
$\lambda_0(\mathbf{r})$, $\lambda_x(\mathbf{r})$,
$\lambda_y(\mathbf{r})$, $\lambda_z(\mathbf{r})$.

The general solution of the problem is too
difficult, if possible, but we can use the
perturbation theory with a small parameter
\begin{equation}
\label{eq:param}
D/J \ll 1 ,
\end{equation}
that is the ratio of typical absolute values of spin-orbit
($D$) and exchange ($J$) interactions. We assume that this
parameter also describes the typical order of magnitude of spatial
derivatives and the spin components responsible for the canting:
\begin{equation}
\label{eq:param2} |\boldsymbol{\nabla}| \sim |u_{\sigma\alpha}|
\sim D/J .
\end{equation}
Another quantity that can be connected with the small value of
canting is $\sqrt{1-m^2}$. In a weak magnetic field, when a spiral
structure still exists, $\sqrt{1-m^2} \sim D/J$. But if the field
is very strong ($g\mu_B H \gg 8 J$ \cite{Dmitrienko2012}), it
induces a ferromagnetic alignment, and $\sqrt{1-m^2} \sim
D/(g\mu_B H) \ll D/J$. In that case we can take the maximal of two
parameters, $D/J$, as constitutive one.

In the next section, we will use the consecutive approximations in
order to find a solution of the system~(\ref{eq:Psi}).

\section{Canting in the first approximation}
\label{sec:1stapprox}

Assuming $\lambda_\alpha^{(0)}=0$, $\alpha=x,y,z$, the zeroth order equations on $D/J$ have the view
\begin{equation}
\label{eq:approx0}
\sum_{n,\{ij\}} \tilde{J}_n (\tau_{i\sigma} + \tau_{j\sigma}) m_\alpha = 0
\end{equation}
and become trivial after the summation on bonds, see Eq.~(\ref{eq:aij}). The first order equations on $D/J$ are
\begin{equation}
\label{eq:approx1a}
\begin{array}{rl}
\sum_{n,\{ij\}} & \{ -\tilde{J}_n (\tau_{i\sigma} - \tau_{j\sigma})  b_{n,ij\mu} \nabla_\mu m_\alpha -\tilde{J}_n (\tau_{i\sigma} \tau_{j\rho} + \tau_{j\sigma} \tau_{i\rho}) u_{\rho\alpha}^{(1)} \\
 & -\varepsilon_{\alpha\beta\gamma} D_{n,ij\beta} (\tau_{i\sigma} - \tau_{j\sigma}) m_\gamma \} + 2 \lambda_0^{(0)} u_{\sigma\alpha}^{(1)} + 2 \lambda_\sigma^{(1)} m_\alpha = 0 ,
\end{array}
\end{equation}
where upper index $(p)$ means that the corresponding term is of the order of $(D/J)^p$. The first summand in curly brackets gives zero in accordance with Eq.~(\ref{eq:x}), two other can be calculated using Eqs.~(\ref{eq:sumtautau}f) and (\ref{eq:tD}). Then,
\begin{equation}
\label{eq:approx1b}
2 (\lambda_0^{(0)} + 4 (\tilde{J}_1 + \tilde{J}_2 + \tilde{J}_3)) u_{\sigma\alpha}^{(1)} + 8 (D_{1+} + D_{2+} + D_{3+}) \varepsilon_{\sigma\alpha\gamma} m_\gamma + 2 \lambda_\sigma^{(1)} m_\alpha = 0 .
\end{equation}
The normalization conditions are
\begin{equation}
\label{eq:norm1a}
u_{\sigma\alpha}^{(1)} u_{\sigma\alpha}^{(1)} = u_{\sigma\alpha} u_{\sigma\alpha} = 1 - m^2 ,
\end{equation}
\begin{equation}
\label{eq:norm1b}
m_\alpha u_{\sigma\alpha}^{(1)} = 0 ,
\end{equation}
where the first equation is of the second order on $D/J$.

Multiplication of Eq.~(\ref{eq:approx1b}) by $m_\alpha$ and summation on $\alpha$ with use of Eq.~(\ref{eq:norm1b}) give
\begin{equation}
\label{eq:lambdas1}
\lambda_\sigma^{(1)} = 0 ,
\end{equation}
therefore
\begin{equation}
\label{eq:u1a}
u_{\sigma\alpha}^{(1)} = - \frac{4 (D_{1+} + D_{2+} + D_{3+}) \varepsilon_{\sigma\alpha\gamma} m_\gamma}{\lambda_0^{(0)} + 4 (\tilde{J}_1 + \tilde{J}_2 + \tilde{J}_3)} .
\end{equation}
The substitution of Eq.~(\ref{eq:u1a}) into Eq.~(\ref{eq:norm1a}) gives
\begin{equation}
\label{eq:lambda01}
\lambda_0^{(0)} + 4 (\tilde{J}_1 + \tilde{J}_2 + \tilde{J}_3) = \frac{4\sqrt{2} |D_{1+} + D_{2+} + D_{3+}| m}{\sqrt{1-m^2}} ,
\end{equation}
and, finally,
\begin{equation}
\label{eq:u1b}
u_{\sigma\alpha}^{(1)} = - sign (D_{1+} + D_{2+} + D_{3+}) \frac{\sqrt{1-m^2}}{\sqrt{2} m} \varepsilon_{\sigma\alpha\gamma} m_\gamma ,
\end{equation}
where the sign of the right parts of Eqs.~(\ref{eq:lambda01}) and (\ref{eq:u1b}) is chosen from the condition of the minimum of the canting energy.

The substitution of Eq.~(\ref{eq:u1b}) into Eq.~(\ref{eq:udef}) gives
\begin{equation}
\label{eq:shat}
\hat{\mathbf{s}}_{i} = \mathbf{m} + \varkappa [\boldsymbol{\tau}_i \times \mathbf{m}] ,
\end{equation}
with
\begin{equation}
\label{eq:kappa}
\varkappa = sign (D_{1+} + D_{2+} + D_{3+}) \frac{\sqrt{1-m^2}}{\sqrt{2} m} .
\end{equation}

From Eqs.~(\ref{eq:approx1b}), (\ref{eq:u1a}) and (\ref{eq:u1b}) we conclude that responsible for the canting is the combination $D_{1+} + D_{2+} + D_{3+}$ of the DM vectors components.

\section{Energy density}
\label{sec:energy}

The contributions to the energy density from the
magnetic moment $\mathbf{m}$ and its
derivatives are
\begin{equation}
\label{eq:calem0}
{\cal E}_m^{(0)} = - \sum_{n,\{ij\}} \tilde{J}_n m_\alpha m_\alpha = -12 (\tilde{J}_1 + \tilde{J}_2 + \tilde{J}_3) m^2 , \\
\end{equation}

\begin{equation}
\label{eq:calem1}
{\cal E}_m^{(1)} = \sum_{n,\{ij\}} \{ -\tilde{J}_n m_\alpha (\mathbf{b}_{n,ij} \cdot \boldsymbol{\nabla}) m_\alpha + \varepsilon_{\alpha\beta\gamma} D_{n,ij\alpha} m_\beta m_\gamma \} = 0 , \\
\end{equation}

\begin{equation}
\label{eq:calem2}
\begin{array}{lll}
{\cal E}_m^{(2)} & = & \sum_{n,\{ij\}} \{ - \frac{1}{2} \tilde{J}_n m_\alpha (\mathbf{b}_{n,ij} \cdot \boldsymbol{\nabla})^2 m_\alpha + \varepsilon_{\alpha\beta\gamma} D_{n,ij\alpha} m_\beta (\mathbf{b}_{n,ij} \cdot \boldsymbol{\nabla}) m_\gamma \\
 & & - (2J_n)^{-1} D_{n,ij\alpha} D_{n,ij\beta} m_\alpha m_\beta \} .
\end{array}
\end{equation}

Using Eqs.~(\ref{eq:sumbb}), (\ref{eq:sumDb2}), and (\ref{eq:DD}) with $x=x_{exch}$ we obtain
\begin{equation}
\label{eq:calem2-2}
{\cal E}_m^{(2)} = - {\cal J} \mathbf{m} \cdot \Delta \mathbf{m} + {\cal D} \mathbf{m} \cdot [\boldsymbol{\nabla} \times \mathbf{m}] - \left( \frac{2D_1^2}{J_1} + \frac{2D_2^2}{J_2} + \frac{2D_3^2}{J_3} \right) m^2 ,
\end{equation}
with
\begin{equation}
\label{eq:calJ}
{\cal J} = \frac{3 \tilde{J}_1^2 + 3 \tilde{J}_2^2 + 3 \tilde{J}_3^2 + 10 \tilde{J}_1 \tilde{J}_2 + 10 \tilde{J}_1 \tilde{J}_3 + 22 \tilde{J}_2 \tilde{J}_3 }{4 (\tilde{J}_1 + \tilde{J}_2 + \tilde{J}_3)} ,
\end{equation}
\begin{equation}
\label{eq:calD}
\begin{array}{ll}
{\cal D} & = - 4 (\mathbf{D}_{1,13} \cdot \mathbf{b}_{1,13} + \mathbf{D}_{2,13} \cdot \mathbf{b}_{2,13} + \mathbf{D}_{3,13} \cdot \mathbf{b}_{3,13}) \\
 & = 8 x_{exch} (D_{1+} + D_{2+} + D_{3+}) - (D_{1+} - D_{1-} + 2 D_{1y}) - (3 D_{2+} + D_{2-} + 2 D_{2y})  - (-D_{3+} + D_{3-} + 2 D_{3y}).
\end{array}
\end{equation}

The first term in Eq.~(\ref{eq:calem2-2}) can be rewritten as
\begin{equation}
\label{eq:efer}
{\cal J} \frac{\partial m_\alpha}{\partial r_\beta} \frac{\partial m_\alpha}{\partial r_\beta} - \frac{1}{2} {\cal J} \boldsymbol{\nabla} \cdot (\boldsymbol{\nabla} m^2) ,
\end{equation}
where the second term gives a contribution to
the surface energy only:
\begin{equation}
\label{eq:esuf}
 - \frac{1}{2} {\cal J} \int_V d\mathbf{r} \boldsymbol{\nabla} \cdot (\boldsymbol{\nabla} m^2) = - \frac{1}{2} {\cal J} \oint_S d\mathbf{f} \cdot \boldsymbol{\nabla} m^2 .
\end{equation}
Far from the transition between paramagnetic
and ferromagnetic states, the absolute value $m$
of the magnetic moment changes slowly, and this
contribution to the energy can be neglected.
However, near the phase transition, $m$ can
undergo considerable changes, and in the crystals
with a significant surface (nanocrystals, thin
films) the term (\ref{eq:esuf}) could play an
important role. In particular, it could be
important for the stabilization of the A-phase
observed in thin films of Fe$_{0.5}$Co$_{0.5}$Si
and FeGe \cite{Yu2010,Yu2011}.

The contributions from the canting into the energy density are
\begin{equation}
\label{eq:caleu1}
{\cal E}_u^{(1)} = - \sum_{n,\{ij\}} \tilde{J}_n (\tau_{i\sigma} + \tau_{j\sigma}) u_{\sigma\alpha}^{(1)} m_\alpha = 0 ,
\end{equation}

\begin{equation}
\label{eq:caleu2}
\begin{array}{lll}
{\cal E}_u^{(2)} & = & \sum_{n,\{ij\}} \{ - \tilde{J}_n (\tau_{i\sigma} - \tau_{j\sigma}) u_{\sigma\alpha}^{(1)} (\mathbf{b}_{n,ij} \cdot \boldsymbol{\nabla}) m_\alpha - \tilde{J}_n (\tau_{i\sigma} + \tau_{j\sigma}) u_{\sigma\alpha}^{(1)} m_\alpha \\
 & & - \tilde{J}_n \tau_{i\sigma} \tau_{j\rho} u_{\sigma\alpha}^{(1)} u_{\rho\alpha}^{(1)} + \varepsilon_{\alpha\beta\gamma} D_{n,ij\alpha} (\tau_{i\sigma} - \tau_{j\sigma}) u_{\sigma\beta}^{(1)} m_\gamma \} \\
 & = & 4 (\tilde{J}_1 + \tilde{J}_2 + \tilde{J}_3) u_{\sigma\alpha}^{(1)} u_{\sigma\alpha}^{(1)} + 8 (D_{1+} + D_{2+} + D_{3+}) \varepsilon_{\sigma\beta\gamma} u_{\sigma\beta}^{(1)} m_\gamma \\
 & = & 4 (\tilde{J}_1 + \tilde{J}_2 + \tilde{J}_3) (1-m^2) - 8 \sqrt{2} |D_{1+} + D_{2+} + D_{3+}| m \sqrt{1-m^2} .
\end{array}
\end{equation}

Thus, we can finally rewrite the bulk energy
density as a function of the magnetic moment
$\mathbf{m}$ accurate within the second order
terms on $D/J$:
\begin{equation}
\label{eq:cale}
\begin{array}{ll}
{\cal E} = & {\cal J} \frac{\partial m_\alpha}{\partial r_\beta} \frac{\partial m_\alpha}{\partial r_\beta} + {\cal D} \mathbf{m} \cdot [\boldsymbol{\nabla} \times \mathbf{m}] + (\tilde{J}_1 + \tilde{J}_2 + \tilde{J}_3) (4 - 16 m^2) \\
 & - 8 \sqrt{2} |D_{1+} + D_{2+} + D_{3+}| m \sqrt{1-m^2} - \left( \frac{2D_1^2}{J_1} + \frac{2D_2^2}{J_2} + \frac{2D_3^2}{J_3} \right) m^2 - 4 g \mu_B \mathbf{H} \cdot \mathbf{m}.
\end{array}
\end{equation}

The first two terms with derivatives are nothing but the
deformation energy of the conventional
phenomenological theory of the chiral magnetics.
However here the values ${\cal J}$ and ${\cal D}$
are expressed through the parameters of the
microscopic theory accordingly to
Eqs.~(\ref{eq:calJ}) and (\ref{eq:calD}). The
following term, on conditions that $m \approx
1$, gives $-12(\tilde{J}_1 + \tilde{J}_2 +
\tilde{J}_3)$, that is the energy of 36
ferromagnetic bonds in first three shells in the
unit cell. Then, the term with $\sqrt{1-m^2}$ is
the contribution of the canting. It is always
negative, which means that the canting is an
important and unavoidable peculiarity of the
magnetic structures of the MnSi-type crystals.
When minimizing the energy on $m$, this term
gives the contribution to the derivative
proportional to $1 / \sqrt{1-m^2}$, which becomes
dominating, when $m \rightarrow 1-0$, impeding
$m$ to exceed 1.

\section{Extra twist induced by canting}
\label{sec:extratwist}

Now we return to Eq.~(\ref{eq:calD}), which defines the DM
parameter $\cal D$ of the phenomenological theory. As it was
mentioned above, from the physical point of view, the DM vectors
are almost perpendicular to the corresponding bonds and,
consequently, the expected value of ${\cal D}$ is small. In
Ref.~\cite{Dmitrienko2012} some speculative estimation has been
performed with use of the well known Keffer rule~\cite{Keffer}
based on the Moriya theory \cite{Moriya60b}, which gives us
following expression for the DM vector
\begin{equation}
\label{eq:keffer}
\mathbf{D}_{12} = D [\mathbf{r}_{1i} \times \mathbf{r}_{2i}] ,
\end{equation}
where $D$ is unknown coefficient, and the
vectors $\mathbf{r}_{1i}$ and $\mathbf{r}_{2i}$
are directed from the positions of 1st and 2nd
magnetic (Mn) atoms to that of an intermediate
nonmagnetic (Si) atom realizing the spin-orbit
interaction. It is evident that $\mathbf{D}_{12}
\perp \mathbf{r}_{12} = \mathbf{r}_{2i} -
\mathbf{r}_{1i}$.

For the Keffer rule one needs the constants $D$ and coordinates
of different intermediate atoms. Surprisingly, a considerable
result can be achieved with a more general rule, namely that the
DM vectors are perpendicular to the bonds. The condition can be
written as
\begin{equation}
\label{eq:Dperpb}
\mathbf{D} \cdot \mathbf{b} (x=x_{real}) = 0 ,
\end{equation}
and, comparing with Eq.~(\ref{eq:calD}), we easily find in this case that
\begin{equation}
\label{eq:calD2}
{\cal D} = 8 (D_{1+} + D_{2+} + D_{3+}) (x_{exch}-x_{real}) ,
\end{equation}
which gives us a new definition of the exchange coordinate
as the manganese atom position inhibiting spiralling when
combining with the Keffer rule. The combination $D_{1+} + D_{2+} +
D_{3+}$ is responsible for the canting in accordance with
Sec.~\ref{sec:1stapprox}, and $x_{exch}$ is a combination of the
exchange interaction constants. Eq.~(\ref{eq:calD2}) can be
interpreted as an evidence of the canting direct
participation in the magnetic structure spiralling. Besides,
the sign of ${\cal D}$ and consequently the magnetic chirality are
determined both by the sign of the canting component $D_{1+} +
D_{2+} + D_{3+}$ of the DM vectors and that of the difference
$x_{exch}-x_{real}$, depending on the exchange constants
$\tilde{J}_1$, $\tilde{J}_2$, and $\tilde{J}_3$.

\begin{figure}[h]
\includegraphics[width=7cm]{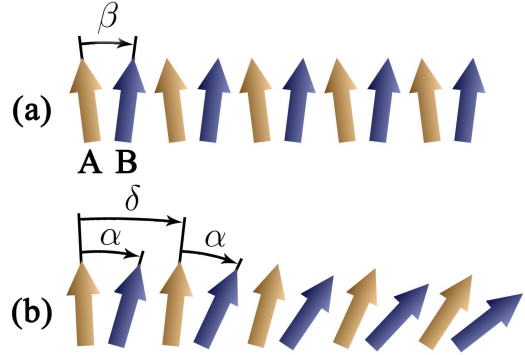}
\caption{\label{fig-chain} (color online). Periodical 1D chain of
magnetic atoms with two spins within a unit cell. The spins can
rotate in a plane, and their directions are fully determined by
the only variable $\varphi$. The spin interaction is described by
the classic Hamiltonian (\ref{eq:Echain}). (a) When $J = 0$, the
chirality is hidden; the lattice is divided onto two sublattices
with the ferromagnetic ordering of spins and the canting angle
$\beta$ between them. (b) When an additional ferromagnetic
interaction with the atoms of the alternate kind is included ($J >
0$), the chirality becomes apparent in the magnetic helix with the
angle $\delta$ between neighboring unit cells; besides, the angle
$\alpha$ between the atoms $A$ and $B$ within a cell remains
the same over the chain.}
\end{figure}

Therefore, the canting, initially considered as a supplementary
microscopic peculiarity of the chiral magnetics
\cite{Chizhikov2012,Dmitrienko2012}, can be in fact an essential
cause of the twisting power. Let us demonstrate by a simple,
albeit not very realistic, example how the canting between
different magnetic sublattices can result in an essential twist
gain. We consider a periodical 1D chain of spins with a local
interaction between them, composed of unit cells containing two
spins, say A and B (Fig.~\ref{fig-chain}). All the spins can
rotate in a plane, and the energy of the chain is a function of
differences of the angles,
\begin{equation}
\label{eq:Echain}
E = \sum_n \left\{ C_R (\varphi_{n}^A-\varphi_{n}^B+\beta)^2 + C_L (\varphi_{n}^A-\varphi_{n-1}^B+\beta)^2 + J (\varphi_{n}^A-\varphi_{n}^B)^2 + J (\varphi_{n}^A-\varphi_{n-1}^B)^2 \right\} .
\end{equation}
Here $\varphi_{n}^A$ and $\varphi_{n}^B$ are orientation angles of the spins A and B of the $n$th cell, laid off from an arbirary direction; $C_R$ and $C_L$ are positive constants. The condition $C_R \neq C_L$ determines the chirality of the structure.

When $J = 0$, Eq.~(\ref{eq:Echain}) describes a periodical magnetic structure with the angle $\beta$ of canting between the spin sublattices A and B (Fig.~\ref{fig-chain}(a)). We suppose that this structure is a result of the competition of two ordering interactions between neighboring spins, a ferromagnetic one and a twisting chiral one; besides, there is a reason, which is not considered here, eliminating the twist in this state.

The question arises: what happens with the structure after
introducing of an additional ferromagnetic ($J > 0$) interaction
with the nearest neighbors from the complementary sublattice? The
minimization of Eq.~(\ref{eq:Echain}) gives the solution in the
helix form
\begin{equation}
\label{eq:phi1phi2}
\begin{array}{l}
\varphi_{n}^A = n \delta , \\
\varphi_{n}^B = n \delta + \alpha ,
\end{array}
\end{equation}
where
\begin{equation}
\label{eq:alpha2}
\alpha = \frac{C_R \beta}{C_R + J}
\end{equation}
is a tilt angle between spins in the unit cell, and
\begin{equation}
\label{eq:ddelta}
\delta = \frac{J (C_R - C_L) \beta}{(C_R + J)(C_L + J)}
\end{equation}
is the twist angle per one period of the chain (Fig.~\ref{fig-chain}(b)).

At the first sight it seems to be paradoxical that the
introducing of an additional aligning interaction induces a twist,
but a simple analysis of the problem shows that there is no
contradiction here. When $|J| \rightarrow \infty$, the angles
$\alpha$ and $\delta$ go to zero as expected. In order to
understand the system behavior for the finite values of $J$,
consider the numerator of Eq.~(\ref{eq:ddelta}). The factor $(C_L
- C_R)$ reflects the degree of the internal chirality of the
structure. The product $J \beta$ reminds of the lever torque, with
$\beta$ playing the role of the lever arm and $J$ being analogous
to the rotating force. We can imagine that the ``aligning force''
$J$, being applied to the initially tilted by the angle $\beta$
sublattices, results in structure distortions, which in their turn
induce the twist due to the potential chirality of the structure
($C_R \neq C_L$). Notice that the change of the sign of the
constant $J$, corresponding to the transition from ferromagnetic
($J > 0$) to antiferromagnetic ($J < 0$) coupling, increases
degree of the twist (reduces the helix pitch) and changes its
handedness.

We expect that the similar effect takes place in the chiral
magnetic MnSi. Indeed, there is an evident similarity of
Eqs.~(\ref{eq:calD2}) and (\ref{eq:ddelta}). In this case the role
of a lever is played by the canting induced by the DM
interactions, whereas the ``force'' is the ferro- or
antiferromagnetic aligning interaction with neighboring atoms.

\section{Estimation of the exchange parameters and the RKKY theory}
\label{sec:rkky}

In analogy with the Dzyaloshinskii--Moriya interaction, the
exchange one is described differently in microscopical and
phenomenological approaches. Although it is evident that exchange
constants of both the theories should be connected to each other,
the connection is found in the nearest neighbors approximation
\cite{Chizhikov2012} to be not so trivial as we could expect.
Indeed, the expression (\ref{eq:calJ}) for ${\cal J}$ in the
approximation of three magnetic shells is strongly different from
the simple proportionality in the former case. However, the
situation seems to be even more intricate, because ${\cal J}$
is not sufficient to induce the ferromagnetic alignment, and
another exchange parameter is also needed, namely
\begin{equation}
\label{eq:Jsum}
J_{\Sigma} = \sum_j \tilde{J}_j ,
\end{equation}
where summation is taken over all the neighbors contributing to
the exchange interaction. Then the condition of the validity of
our approximation can be written as two inequalities,
\begin{equation}
\label{eq:Jsum0}
J_{\Sigma} > 0 ,
\end{equation}
\begin{equation}
\label{eq:Jcal0}
{\cal J} > 0 ,
\end{equation}
both equally important. Here $J_{\Sigma}$ is nothing but taken
with an opposite sign energy of an individual spin interaction
with its magnetic surroundings in the untwisted (ferromagnetic)
state, and Eq.~(\ref{eq:Jsum0}) guarantees stability of the state
relative to the change of the spin sign. Eq.~(\ref{eq:Jcal0}) in
its turn guarantees stability relative to the spin small
rotations, providing smallness of the magnetic moment gradients.
Therefore, only combined use of the conditions leads to a
ferromagnetic ordering with a weak spiralling.

Neglecting the contribution of the DM interaction into symmetrical
exchange (i.e. using $J_n$ instead of $\tilde{J}_n$), we can
estimate $J_\Sigma$ and ${\cal J}$ in the frame of the RKKY theory
\cite{Ruderman1954,Kasuya1956,Yosida1957,VanVleck1962}, which is
applicable to the itinerant magnetics. Indeed, in this model
$J_n$ is a simple function of the distance between interacting
atoms,
\begin{equation}
\label{eq:Jrkky}
J_n \equiv J(b_n) \sim -F(2 k_F a b_n) ,
\end{equation}
\begin{equation}
\label{eq:F}
F(x) = \frac{x \cos x - \sin x}{x^4} .
\end{equation}
Here $k_F$ is the Fermi wave number, $b_n$
is the dimensionless distance to the $n$th magnetic shell.

It follows from Eqs.~(\ref{eq:Jrkky}), (\ref{eq:F}) that atoms
situated at approximately the same distances, e.g. $b_2$ and
$b_3$, make similar contributions to the exchange energy.
Therefore, together with the 2nd and 3rd shells it is necessary
to take into account the 4th shell corresponding to the atoms
separated by lattice periods. Because the atoms of the 4th
shell belong to the same magnetic sublattice, they do not affect
the canting between different sublattices, i.e. they leave
$u_{\sigma\alpha}$ and $x_{exch}$ unchanged and give simple
additive contributions to $J_\Sigma$ and ${\cal J}$.

All the magnetic shells have 6 atoms, so
\begin{equation}
\label{eq:Jsum-2}
J_{\Sigma} = 6 (J_1 + J_2 + J_3 + J_4) .
\end{equation}
In order to calculate the additive from the 4th shell to ${\cal
J}$, we can write, for example, the interaction energy of two
spins separated by the period $(1,0,0)$ of the lattice,
\begin{equation}
\label{eq:J4part}
-J_4 \mathbf{s}(\mathbf{r}) \cdot \mathbf{s}(\mathbf{r}+(1,0,0)) \approx -J_4 \left( 1 + \mathbf{s} \cdot \frac{\partial\mathbf{s}}{\partial x} + \frac12 \mathbf{s} \cdot \frac{\partial^2\mathbf{s}}{\partial x^2} \right) .
\end{equation}
If we take the sum over 6 bonds, multiply it by the number of
magnetic atoms in the unit cell and take into account that all the
bonds in that sum are taken twice, then the correction to the
energy density can be written as
\begin{equation}
\label{eq:J4toE}
\Delta{\cal E} = -12 J_4 - 2 J_4 \mathbf{m} \cdot \Delta\mathbf{m} .
\end{equation}
It is obvious from the comparison with Eq.~(\ref{eq:calem2-2})
that the additive to ${\cal J}$ from the 4th magnetic shell is $2
J_4$.

\begin{figure}[h]
\includegraphics[width=7cm]{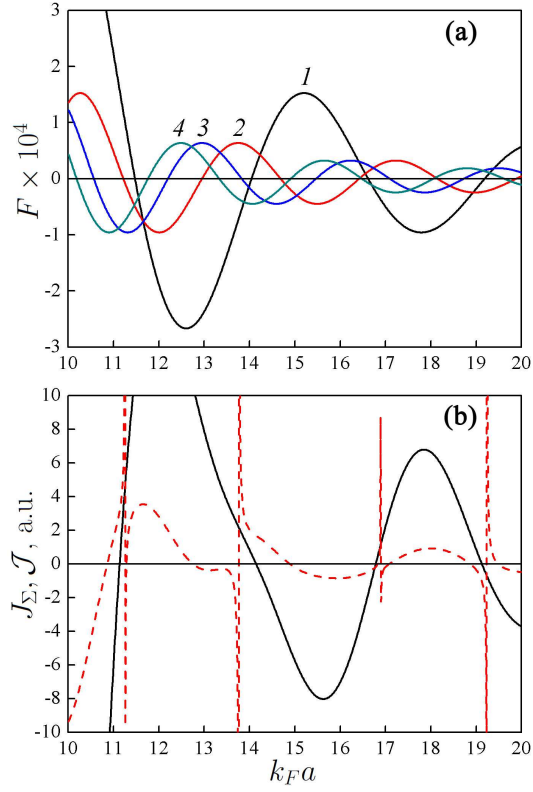}
\caption{\label{fig-J} (color online). (a) The $F_n \equiv F(2 k_F
a b_n)$ dependences on the Fermi wave number for four magnetic
shells. The distances $b_1$--$b_4$ are chosen as for MnSi. The
minima of $F_1$ give the areas with $J_\Sigma, {\cal J} > 0$. (b)
The $J_\Sigma$ (solid line) and ${\cal J}$ (dash line) dependences
on the Fermi wave number. The plots show strong oscillations.
The jumps of ${\cal J}$ correspond to the zeros of $J_1+J_2+J_3$.
The areas with $J_\Sigma, {\cal J} > 0$ determine weakly twisted
ferromagnetic states. The most close to the known for MnSi value
$k_F a = 16.4$ plateau is in the area $17.1 < k_F a < 18.9$.}
\end{figure}

The parameters $F_n \equiv F(2 k_F a b_n)$ are oscillating
functions of the argument $k_F a$, which are easy to calculate
using real distances between magnetic atoms. In MnSi for
$x=0.138$ we find $b_1=0.613$, $b_2=0.908$, $b_3=0.964$ and
$b_4=1$. Fig.~\ref{fig-J}(a) shows the functions $F_1$--$F_4$
plotted in the area $10 < k_F a < 20$, where the value $k_F a =
16.4$ for MnSi is situated ($k_F = 3.6$\AA$^{-1}$
\cite{Kirkpatrick2009}, $a = 4.56$\AA). The graphs show that
Eqs.~(\ref{eq:Jsum0}), (\ref{eq:Jcal0}) can be satisfied together
in the areas of negative values of $F_1$. Fig.~\ref{fig-J}(b)
represents dependences of $J_\Sigma$ and ${\cal J}$ on $k_F a$,
calculated for the same area. Both $J_\Sigma$ and ${\cal J}$ are
oscillating alternating-sign functions; the jumps of ${\cal J}$
correspond to the zeros of $J_1 + J_2 + J_3$. The nearest to the
value $k_F a = 16.4$ ``plateau'' with positive $J_\Sigma$ and
${\cal J}$ is in the area $17.1 < k_F a < 18.9$.

Notice that the behavior of ${\cal J}$ at some $k_Fa$ could seem
paradoxical. For example, the divergence of ${\cal J}
\rightarrow +\infty$ would mean a strong suppression of
long-wavelength fluctuations. However, the paradox can be
resolved by taking into account the behavior of $J_\Sigma$.
Indeed, if $J_1 + J_2 + J_3 = 0$, then $J_\Sigma \sim J_4$, where
$J_4$ is nothing but the coupling constant of the spins belonging
to the same sublattice. This means that, when ${\cal J}
\rightarrow +\infty$, the connection between sublattices gets
broken. Thus, when the spins of three sublattices are aligned in
the same direction, the spins of the fourth one can have an
arbitrary direction, even if the condition $J_4 > 0$ guarantees
the ferromagnetic order within the sublattice. The foregoing means
that in addition to Eqs.~(\ref{eq:Jsum0}), (\ref{eq:Jcal0}), we
should introduce another inequality,
\begin{equation}
\label{eq:Jsum0-2}
J_{\Sigma}^\prime = \left. \sum_j \right.^\prime \tilde{J}_j > 0 ,
\end{equation}
determining the ferromagnetic connection of the four magnetic sublattices. Here the sum is taken over the bonds connecting atoms belonging to different sublattices. In the approximation of four magnetic shells, $J_{\Sigma}^\prime = 6 (J_1 + J_2 + J_3) = J_{\Sigma} - 6 J_4$, therefore the zeros of $J_{\Sigma}^\prime$ coincide with the jumps of ${\cal J}$.

\begin{figure}[h]
\includegraphics[width=7cm]{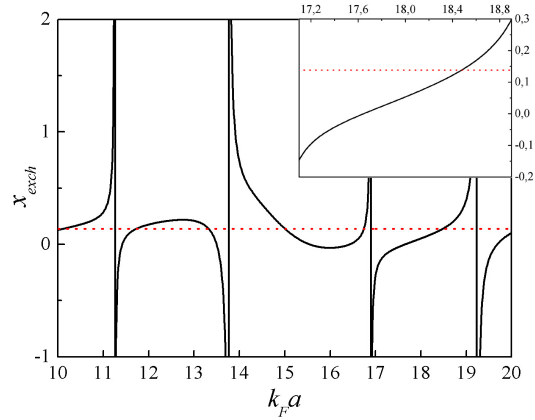}
\caption{\label{fig-xexch} (color online). Exchange coordinate
$x_{exch}$ calculated in the RKKY model. The jumps of the function
correspond to the zeros of $J_1+J_2+J_3$. The dot line identifies
the real value $x_{real}=0.138$ for MnSi. The inset shows the
$x_{exch}$ dependence in the corresponding to positive $J_\Sigma$
and ${\cal J}$ area $17.1 < k_F a < 18.9$, where $-0.28 < x_{exch}
- x_{real} < 0.16$. The chirality of the magnetic structure change
its sign, when $x_{exch} = x_{real}$.}
\end{figure}

The RKKY model allows as well to estimate the exchange coordinate
$x_{exch}$. Because, according to Eq.~(\ref{eq:calD2}), the degree
of the twist is determined by the difference $x_{exch} -
x_{real}$, it is useful to have an idea about how much this
difference could be. In the nearest neighbors approximation
$x_{exch} = \frac18$, which is close to the real value $x_{real} =
0.138$ for MnSi. However, when taking into account the
contributions from the 2nd and 3rd shells, $x_{exch}$ can have
arbitrary large positive and negative values near the zeros of
$J_1 + J_2 + J_3$ (Fig.~\ref{fig-xexch}). Close to the minima of
$F_1$ $J_2$ and $J_3$ are small in comparison with $J_1$ and,
therefore, $x_{exch} \approx \frac18$, see inset in
Fig.~\ref{fig-xexch}. Notice that the zeros of $J_1 + J_2 + J_3$
do not result in a divergence of the wave number $k={\cal
D}/2{\cal J}$, because ${\cal J}$ in the denominator and
$x_{exch}$ in the numerator increase simultaneously.

As it is seen from Fig.~\ref{fig-xexch} and the inset in it, the
difference $x_{exch} - x_{real}$ can change its sign depending on
$k_F$. It gives a possibility to control the magnetic structure
chirality by varying the concentration of different elements in
the crystal.

Eqs.~(\ref{eq:Jsum0}), (\ref{eq:Jcal0}) are evident preconditions
of the experimentally observed ferromagnetic order in MnSi.
Nevertheless, we can not preclude that one of the constants
$J_\Sigma$, ${\cal J}$, or both these parameters can have negative
values. For example, the condition $J_\Sigma < 0$ does not surely
result in an antiferromagnetic order. The strong frustrations
intrinsic for the system, e.g. the triangles of bonds, and
nonsymmetric DM interactions can induce a small magnetic moment
and lead to a ferri- or a weak ferromagnetic order. It could
explain the weak magnetic moment observed in MnSi ($g \approx
0.4$). When ${\cal J} < 0$, the contributions to the energy
density with higher spatial derivatives should be taken into
account, which can stabilize the helix pitch.

\section{Canting and magnetic diffraction}
\label{sec:experimental}

In Sec.~\ref{sec:1stapprox}, we obtain Eq.~(\ref{eq:shat}), which
says that in the first approximation the canting can be described
as spin rotations by the same small angle around corresponding
3-fold axes. A similar expression for the ferromagnetic state
caused by a strong magnetic field  was found in
Ref.~\cite{Dmitrienko2012}, where an approach had been suggested
to measure the canting using neutron or X-ray magnetic
diffraction. In Ref.~\cite{Dmitrienko2012}, the angle of spin tilt
was proportional to $\delta = D_{1+} / 4 J_1$. In order to obtain
that result we can find the coefficient $\varkappa \sim
\sqrt{1-m^2} / \sqrt{2} m$ for untwisted state. The minimization
of Eq.~(\ref{eq:cale}) on magnetic moment modulus $m$, assuming
that $\mathbf{H} \parallel \mathbf{m} = const$, gives
\begin{equation}
\label{eq:kappa2}
\varkappa = \frac{D_{1+} + D_{2+} + D_{3+}}{4 (J_1 + J_2 + J_3 + g \mu_B H / 8)}
\end{equation}
Notice that  $\varkappa =\delta$, when $D_2 = D_3 = J_2 = J_3 = H
= 0$. The corresponding expressions for the structure
factors of purely magnetic reflections $00\ell(\ell=2n+1)$ can be
easily found from Ref. \cite{Dmitrienko2012}.

Notice that there is another contribution to the reflections,
induced by the anisotropy of magnetic susceptibility tensor
\cite{Gukasov2002,Cao2009}. The contributions can be
distinguished, because (i) the tilts have different directions,
(ii) the canting effect does not depend on the magnetic field
modulus, whereas the tilts induced by the susceptibility tensor
anisotropy are proportional to $H$.

The fact that both the twist and canting are determined by the
same DM vectors components gives an additional possibility of
numerical verification of the theory. Indeed, excluding $D_{1+} +
D_{2+} + D_{3+}$ from Eqs.~(\ref{eq:kappa2}) and (\ref{eq:calD2}),
we can connect the canting angle in unwound state and the wave
number $k = {\cal D} / 2 {\cal J}$ of magnetic helices:
\begin{equation}
\label{eq:kappa3}
\varkappa = \frac{\cal J}{16 (J_1 + J_2 + J_3 + g \mu_B H / 8) (x_{exch} - x_{real})} k .
\end{equation}
Excepting observable physical values, Eq.~(\ref{eq:kappa3})
contains only exchange interaction constants, which are easier to
calculate using {\it ab initio} methods than the DM vectors.

\section{Summary and discussion}
\label{sec:discussion}

An essential difference of the microscopic description of the
magnetic properties of the MnSi-type crystals from the
phenomenological one is the usage of the pseudovector $\mathbf{D}$
instead of pseudoscalar ${\cal D}$ when describing the
Dzyaloshinskii--Moriya interaction. The presence of the extra
parameters (pseudovector components) results in the existence of a
local canting, the feature not studied yet in magnetic twisted
structures. An important problem solved in the present work is how
to distinguish the components of $\mathbf{D}$ responsible for the
twist and the canting. Hopkinson and Kee in
Ref.~\cite{Hopkinson2009} showed numerically (in the nearest
neighbors approximation) that responsible for the twist were the
components of the $\mathbf{D}$-vectors lying along the bonds,
whereas for the canting did the $\mathbf{D}$ components directed
perpendicular to the bonds and lying in the planes of bond
triangles in the trillium lattice. In Ref.~\cite{Chizhikov2012} we
specified the result and showed that real components of
$\mathbf{D}$ inducing the twist and the canting lay along
crystallographic directions closed to those found in
Ref.~\cite{Hopkinson2009}. Nevertheless, the problem remained that
in accordance with the quantum mechanical description the DM
vectors should be perpendicular or almost perpendicular to the
bonds. In other words, the $\mathbf{D}$ components responsible for
the twist accordingly to Refs.~\cite{Hopkinson2009,Chizhikov2012}
could be diminutive. In order to solve the problem we take into
account the contribution of non-nearest neighbors in the magnetic
interaction.  Surprisingly, it appears that in the case, when all
the DM vectors are perpendicular to the bonds, the spiralling is
determined by the same $\mathbf{D}$ components as the canting. It
leads to the conclusion that the canting, initially being
considered as an additional microscopic effect in relation to the
global twist, can in fact serve as a cause of the abnormal twist
experimentally found in the MnSi-type crystals, particularly in
MnGe. It is also important that the contribution of non-nearest
magnetic neighbors should be taken into account.

In the simplest phenomenological theory describing twisted
magnetic structures, it is supposed that $|\mathbf{M}| = const$,
which is roughly true at low temperatures, far below the phase
transition from the paramagnetic state. However, this condition
makes energetically unfavorable such structures as the Skyrmions
and their lattices, associated with the A-phase observed close by
the transition point. In order to overcome this problem as well as
to describe the critical phenomena, two additional terms, $M^2$
and $M^4$, limiting the value of magnetic moment are included into
the free energy \cite{Bak80}. The presence of the terms decreases
$M$ in the regions with a large density of the magnetic energy,
thereby decreasing the energy of the whole structure. Thus, in
Ref.~\cite{Roessler2011} the terms $M^2$ and $M^4$ are used in
order to calculate the energy of Skyrmions. It is found that $M$
decreases considerably nearby the core of the Skyrmion, where the
magnetic moment $\mathbf{M}$ has the opposite direction to the
external magnetic field, and has a maximum at some distance from
the core, where the energy gain from the double twist is maximal.
However, in the latter case the magnetic moment exceeds its
saturation value $M_0$, which is not acceptable for physical
reasons. In the present work we show that there should be a
contribution from the canting $\sim m \sqrt{1 - m^2}$ into the
energy density, which can play the same role as $M^2$ and $M^4$,
but does not allow the magnetic moment modulus to exceed the
saturation threshold. Besides the canting, the thermal
fluctuations of spins also contribute to the reduction of the
magnetic moment $M$. Nearby the transition point the amplitude of
the fluctuations can be comparable with the canting. Moreover,
the less is the effective local field $\mathbf{h}_{eff,i} =
-\partial E / \partial \mathbf{s}_i$, acting on the individual
spin $\mathbf{s}_i$, the more are fluctuations and cantings.
Therefore, the reinforcement both of the thermal fluctuations
and the canting have the same cause, so they should give a similar
effect.

If the canting would be observed in MnSi, the direct confirmation
of non-nearest neighbors effect could be possible using
Eq.~(\ref{eq:kappa2}), connecting the propagation number of the
magnetic helices with the magnitude of the residual canting in the
unwound state in a strong magnetic field. The equation involves
only the exchange constants $J_n$ and does not depend on the DM
vectors. The possibility of an experimental proof of the theory
should stimulate {\it ab intio} calculations of the interaction
constants. Some semi-quantitative estimations are made in the
present work with use of the RKKY model. However problems still
remain. For example, the RKKY parameter $k_F a = 16.4$ for MnSi
corresponds to the area, where $J_\Sigma < 0$ and $ {\cal J} <
0$. More realistic calculations would give a tip about the
direction of the further search.

It has been found in Ref.~\cite{Grigoriev2009} that the
propagation number $k$ of the magnetic helix in the alloy
Fe$_{1-x}$Co$_x$Si is strongly dependent on the concentration $x$
of the cobalt atoms. Thus, when $x$ changes from 0.05 to 0.15, the
helix period decreases abruptly in several times. It can be
explained, in particular, by the strong dependence of the Fermi
wave number $k_F$ on the cobalt impurity concentration, which
also effects on the exchange parameters $J_\Sigma$, ${\cal J}$
and $x_{exch}$. This phenomenon could be also responsible for
the recently observed \cite{GrigorievFKS-2013} change of the sign
of the magnetic chirality in Mn$_{1-x}$Fe$_{x}$Ge alloys. This is
drastically different from the usually supposed change of the sign
of the DM interaction. Indeed, Eq.~(\ref{eq:calD2}) shows that
the chirality can change even if the microscopic DM interaction,
defined by the vectors $\mathbf{D}$, remains constant. In this
case the sign change is due to the interplay between the
exchange parameters $J_1$, $J_2$, and $J_3$ of three magnetic
shells. Thus a potential possibility arises to control the
sign of the magnetic chirality by varying the concentration of the
different components and therefore affecting the Fermi wave
number $k_F$. Notice that the possibility of such effect
becomes evident only when the interactions with non-nearest
neighbors are taken into account.

Another interesting fact, not being yet explained within the
framework of microscopic theories, is the helix ordering along
some special directions, e.g. $\langle 111 \rangle$ or $\langle
100 \rangle$. Usually this ordering is associated with a weak
anisotropic exchange \cite{Bak80}, but in fact the ordinary DM
interaction also can result in the appearing of cubic anisotropic
terms in the energy of spiral orientation in the lattice with a
cubic space group. These contributions of the order $(D/J)^4$ will
determine the critical magnetic field $H_{c1}$, at which the helix
comes off from its preferable zero-field direction. This is in a
good agreement with the observed ratio of the first and second
critical fields $H_{c1} \ll H_{c2}$, because it follows from
our estimations that $H_{c1} / H_{c2} \sim (D/J)^{2}$. Indeed,
e.g. the period of the magnetic helix in MnSi makes about 40 unit
cell parameters, which gives the value $2\pi / 40$ for the
propagation number modulus $|k|$. On the other part, it follows
from the phenomenological description that $k = {\cal D} / (2
{\cal J})$, or $D/J \sim \pi / 10 \approx 0.3$. This gives us the
estimation $H_{c1} / H_{c2} \sim 0.1$, which is in a good
agreement with the experimental data. In our previous work
\cite{Chizhikov2012} we proposed a coarse grain approximation,
which allowed us to calculate only the contributions to the
energy proportional to $(D/J)^2$. In the present work a new
approach has been developed permitting more precise calculations
of the energy. In particular, the terms of the order of $(D/J)^4$
would give us a contribution of the Dzyaloshinskii--Moriya
interaction into the energy of the cubic anisotropy responsible
for (i) the ordering of the spiral axes along selected
crystallographic directions, e.g. $\langle 111 \rangle$ in MnSi,
in the absence of external magnetic field; (ii) the orientation of
the triangle Skyrmion lattice in the A-phase observed in these
crystals \cite{Grigoriev2006a,Munzer2010,Adams2011}.

\section*{Acknowledgements}

We are grateful to M.~V. Gorkunov, S.~V. Demishev, S.~V. Maleyev,
S.~V. Grigoriev, V.~A. Dyadkin and S.~S. Doudoukine for useful
discussions and encouragement. This work was supported by two
projects of the Presidium of the Russian Academy of Sciences:
``Matter at high energy densities; Substance under high static
compression'' and ``Diffraction of synchrotron radiation in
multiferroics and chiral magnetics''.

\appendix
\section{}
\label{appendixA}

The vectors $\boldsymbol{\tau}_i$, $\boldsymbol{\tau}_j$, $(\boldsymbol{\tau}_i - \boldsymbol{\tau}_j)$, $(\boldsymbol{\tau}_i + \boldsymbol{\tau}_j)$, $[\boldsymbol{\tau}_i \times \boldsymbol{\tau}_j]$, $\mathbf{b}_{n,ij}$, $\mathbf{D}_{n,ij}$ ($n = 1, 2, 3$) change with use of the same symmetry transformations of the point group $23$, when the index $ij$ passes trough 12 possible values $\{12, 13, 14, 21, 23, 24, 31, 32, 34, 41, 42, 43\}$. Therefore we can use the formulae
\begin{equation}
\label{eq:aij}
\sum_{\{ij\}} \mathbf{a}_{ij} = 0 ,
\end{equation}

\begin{equation}
\label{eq:aijbij}
\sum_{\{ij\}} a_{ij\alpha} b_{ij\beta} = 4 (\mathbf{a}_{12} \cdot \mathbf{b}_{12}) \delta_{\alpha\beta} ,
\end{equation}
where $\mathbf{a}_{ij}$ and $\mathbf{b}_{ij}$ are the vectors from the above-listed set. Thereby the following summations can be easily made:
\begin{subequations}     % need amsmath package
\label{eq:sumtautau}
\begin{eqnarray}
\sum_{\{ij\}} (\boldsymbol{\tau}_i - \boldsymbol{\tau}_j)_{\alpha} (\boldsymbol{\tau}_i + \boldsymbol{\tau}_j)_{\beta} = 0 , \\
\sum_{\{ij\}} (\boldsymbol{\tau}_i \pm \boldsymbol{\tau}_j)_{\alpha} [\boldsymbol{\tau}_i \times \boldsymbol{\tau}_j]_{\beta} = 0 , \\
\sum_{\{ij\}} (\boldsymbol{\tau}_i - \boldsymbol{\tau}_j)_{\alpha} (\boldsymbol{\tau}_i - \boldsymbol{\tau}_j)_{\beta} = 32 \delta_{\alpha\beta} , \\
\sum_{\{ij\}} (\boldsymbol{\tau}_i + \boldsymbol{\tau}_j)_{\alpha} (\boldsymbol{\tau}_i + \boldsymbol{\tau}_j)_{\beta} = 16 \delta_{\alpha\beta} , \\
\sum_{\{ij\}} [\boldsymbol{\tau}_i \times \boldsymbol{\tau}_j]_{\alpha} [\boldsymbol{\tau}_i \times \boldsymbol{\tau}_j]_{\beta} = 32 \delta_{\alpha\beta} , \\
\sum_{\{ij\}} (\tau_{i\alpha} \tau_{j\beta} + \tau_{j\alpha} \tau_{i\beta}) = -8\delta_{\alpha\beta} .
\end{eqnarray}
\end{subequations}

Using Eqs.~(\ref{eq:sumtautau}a)-(\ref{eq:sumtautau}e) we can calculate the sums containing products of the vectors $\mathbf{b}_{n,ij}$, $\mathbf{D}_{n,ij}$ ($n = 1, 2, 3$):
\begin{subequations}     % need amsmath package
\label{eq:sumtaub}
\begin{eqnarray}
\sum_{\{ij\}} (\boldsymbol{\tau}_i - \boldsymbol{\tau}_j)_{\alpha} b_{1,ij\beta} = 4 (-8x+1) \delta_{\alpha\beta} , \\
\sum_{\{ij\}} (\boldsymbol{\tau}_i - \boldsymbol{\tau}_j)_{\alpha} b_{2,ij\beta} = 4 (-8x+3) \delta_{\alpha\beta} , \\
\sum_{\{ij\}} (\boldsymbol{\tau}_i - \boldsymbol{\tau}_j)_{\alpha} b_{3,ij\beta} = 4 (-8x-1) \delta_{\alpha\beta} ,
\end{eqnarray}
\end{subequations}

\begin{equation}
\label{eq:tD}
\sum_{\{ij\}} (\boldsymbol{\tau}_i - \boldsymbol{\tau}_j)_{\alpha} D_{n,ij\beta} = 8D_{n+}\delta_{\alpha\beta} ,
\end{equation}

\begin{equation}
\label{eq:DD}
\sum_{\{ij\}} D_{n,ij\alpha} D_{n,ij\beta} = 4 D_n^2 \delta_{\alpha\beta} ,
\end{equation}

\begin{subequations}     % need amsmath package
\label{eq:sumbb}
\begin{eqnarray}
\sum_{\{ij\}} b_{1,ij\alpha} b_{1,ij\beta} = (32 x^2 - 8 x + 2) \delta_{\alpha\beta} , \\
\sum_{\{ij\}} b_{2,ij\alpha} b_{2,ij\beta} = (32 x^2 - 24 x + 6) \delta_{\alpha\beta} , \\
\sum_{\{ij\}} b_{3,ij\alpha} b_{3,ij\beta} = (32 x^2 + 8 x + 2) \delta_{\alpha\beta} ,
\end{eqnarray}
\end{subequations}

\begin{subequations}     % need amsmath package
\label{eq:sumDb2}
\begin{eqnarray}
\sum_{\{ij\}} D_{1,ij\alpha} b_{1,ij\beta} = ((-8x+1) D_{1+} - D_{1-} + 2 D_{1y}) \delta_{\alpha\beta} , \\
\sum_{\{ij\}} D_{2,ij\alpha} b_{2,ij\beta} = ((-8x+3) D_{2+} + D_{2-} + 2 D_{2y}) \delta_{\alpha\beta} , \\
\sum_{\{ij\}} D_{3,ij\alpha} b_{3,ij\beta} = ((-8x-1) D_{3+} + D_{3-} + 2 D_{3y}) \delta_{\alpha\beta} .
\end{eqnarray}
\end{subequations}

\newpage
\section*{Figures}

\end{document}